\documentclass[twocolumn,trackchanges]{aastex631}

\usepackage{multirow}
\usepackage[normalem]{ulem}
\usepackage{multirow}
\usepackage{soul}
\newcommand\blfootnote[1]{%
  \begingroup
  \renewcommand\thefootnote{}\footnote{#1}%
  \addtocounter{footnote}{-1}%
  \endgroup
}

\begin{document}
 
\title{Imprint of “Local Opacity” Effect in Gamma-Ray Spectrum of Blazar Jet}

\author[0000-0001-5507-7660]{Sushmita Agarwal}
\author[0000-0002-5656-2657]{Amit Shukla}
\affiliation{ Department of Astronomy, Astrophysics and Space Engineering, Indian Institute of Technology Indore, Khandwa Road, Simrol, Indore, 453552, India}

\author[0000-0002-2950-6641]{Karl Mannheim}
\affiliation{Julius-Maximilians-Universität Würzburg, Fakultät für Physik und 
Astronomie, Institut für Theoretische Physik und Astrophysik, 
Lehrstuhl für Astronomie, Emil-Fischer Str. 31, D-97074 Würzburg, 
Germany}

\author[0000-0001-5424-0059]{Bhargav Vaidya}
\affiliation{ Department of Astronomy, Astrophysics and Space Engineering, Indian Institute of Technology Indore, Khandwa Road, Simrol, Indore, 453552, India} 

\author[0000-0002-8008-2485]{Biswajit Banerjee}
\affiliation{ Gran Sasso Science Institute, Viale F. Crispi 7, L’Aquila (AQ), I-67100, Italy}
\affiliation{ INFN - Laboratori Nazionali del Gran Sasso, L'Aquila (AQ), I-67100, Italy}

\begin{abstract}
Relativistic jets from accreting supermassive black holes at cosmological distances can be powerful emitters of $\gamma$-rays. However, the precise mechanisms and locations responsible for the dissipation of energy within these jets, leading to observable $\gamma$-ray radiation, remain elusive. We detect evidence for an intrinsic absorption feature in the $\gamma$-ray spectrum at energies exceeding $10\,$GeV, presumably due to the photon$-$photon pair production of $\gamma$-rays with low-ionization lines at the outer edge of broad-line region (BLR), during the high-flux state of the flat-spectrum radio quasar PKS~1424$-$418. The feature can be discriminated from the turnover at higher energies resulting from $\gamma$-ray absorption in the extragalactic background light.
It is absent in the low-flux states, supporting the interpretation that powerful dissipation events within or at the edge of the BLR evolve into fainter $\gamma$-ray emitting zones outside the BLR, possibly associated with the moving very long baseline interferometry radio knots. The inferred location of the $\gamma$-ray emission zone is consistent with the observed variability time scale of the brightest flare, provided that the flare is attributed to external Compton scattering with BLR photons.
\end{abstract}

\keywords{Blazars (164); Gamma-ray astronomy (628); Relativistic jets (1390);  Galactic and extragalactic astronomy(563); Flat-spectrum radio quasars(2163); High energy astrophysics(739)}

\vspace{-1em}
\section{Introduction} \label{sec:intro}
\blfootnote{Corresponding author: Sushmita Agarwal}
\blfootnote{Email: \href{sush.agarwal16@gmail.com}{sush.agarwal16@gmail.com}}
Accreting supermassive black holes spew out collimated,  relativistic jets. Specifically when aligned with our line of sight, the emission from the jets is Doppler boosted, thus rendering them visible up to high redshifts. In a leptonic scenario, blazars---aligned jetted objects---emit high-energy radiation through the upscattering of soft seed photons.  These soft photons could originate within the jet through the synchrotron-self-Compton process or could be contributed by an external radiation field \citep{2009MNRAS.397..985G}. 

Particularly in flat spectrum radio quasars (FSRQs), the observed high Compton dominance implies an increased influence of external seed photons from the broad-line region (BLR ; \citealt{2009MNRAS.397..985G}). Typically FSRQs are characterized by high black hole masses, enhanced radiative efficiency from accretion disks, and accretion rates near the Eddington limit. As a result, they possess a luminous disk \citep{Maraschi_2003}. Part of disk radiation is reprocessed, contributing to the luminosity of the BLR and torus, thereby influencing the external jet environment close to the black hole \citep{2011MNRAS.411..901G,10.1111/j.1365-2966.2012.20442.x}. $\gamma$-ray photons produced near the black hole on propagating through this dense photon environment, are expected to leave an imprint of photon$-$photon pair creation at energies from 10 to 200 GeV \citep{Liu_2006}. Thus, an expected observational signature of production of high-energy photons within BLR, occurring within parsec scales from central engine, would manifest as a cutoff feature at $E>10\,$GeV in high-energy spectrum. However, the absence of such BLR-induced cutoff in high-energy spectrum of FSRQs raises concerns about the Compton dominance typically observed in such sources \citep{10.1093/mnras/sty887}. The absence of cutoff above 10 GeV indicates that the emission region of high-energy photons is outside the influence of the BLR. Multiple detection of very high-energy TeV photons from FSRQs such as 3C~279 (\citealt{2011A&A...530A...4A}), PKS~1510$-$089 (\citealt{2013A&A...554A.107H}, \citealt{2018A&A...619A.159M}), PKS~1222$+$216 (\citealt{2011ApJ...730L...8A}), and PKS~1441$+$25 (\citealt{2015ApJ...815L..22A}), further supports the possibility of high-energy emission region beyond the BLR  \citep{Liu_2006, 10.1143/PTPS.151.186}.  In contrast to the previous results, \citet{Poutanen2010} presented evidence of a break in the high-energy spectrum. The idea gains further support from Fermi-LAT observation of 3C~454.3  and 4C~+21.35 \citep{2013Leon, 2013Isler,2011ApJ...733...19T, Stern_Poutanen_2014}. This constrains the emission site within the region of influence of the BLR. Additionally, the detection of fast variability typically observed in FSRQs implies the production of high-energy $\gamma$-rays towards the edge of the BLR via a magnetic reconnection scenario  \citep{Shukla2020,2023Agarwal}.

In this context, we studied a high-redshift ($z=1.522$) FSRQ PKS~1424$-$418 through flux-resolved spectroscopy. The source recently exhibited exceptional outbursts during 2022, reaching 10 times the average flux level. The large black hole mass of $4.5\times10^{9}\,M_{\odot}$ in PKS~1424$-$418 provides a possibility of strong accretion rate and thus sufficient seed photons from BLR for the observed Compton dominance ($q\sim30$; \citet{2021MNRAS.501.2504A}). The structure of the paper is as follows: Section \ref{sec:method} discusses the methods and techniques, Section \ref{sec:result} presents the results, and a discussion is provided in Section \ref{sec:discussion}. Our results are summarized in Section \ref{sec:summary}.
\section{Methods and Techniques} \label{sec:method}
\subsection{Data Analysis: Fermi-LAT} \label{subsec:Fermi_analysis}
To examine the varying flux states of PKS~1424$-$418, we have chosen  $\approx$15 yr of $\gamma$-ray data within the energy range of $0.1-300$ GeV. This data were recorded by the Large Area Telescope (LAT) on board the Fermi Gamma-ray Space Telescope, spanning from 2008 August 4, to 2023 March 21 \citep{2009ApJ...697.1071A}. Standard analysis procedure\footnote{\url{https://fermi.gsfc.nasa.gov/ssc/data/analysis/documentation/Cicerone/Cicerone_Data_Exploration/Data_preparation.html}} provided by Fermi Science tools and the open-source Fermipy package \citep{Wood:2017TJ} are used to analyze data for the source in the energy range between $0.1-300$ GeV using the latest instrument response function {\tt\string P8R3\_SOURCE\_V3}.  All the photons coming within $20^\circ$ of the source location have been analyzed to account for the broad PSF of the telescope. Additionally, a zenith angle cut of $90^\circ$, {\tt\string GTMKTIME } cut of {\tt\string DATA\_QUAL} $>$ 0 \&\& {\tt\string LAT\_CONFIG==1} and evtype=3 were used in the analysis. Only those events are considered for analysis that is highly probable of being photons by applying a {\tt\string GTSELECT} cut on the event class to account for the {\tt\string SOURCE } class event using evclass=128. 
Spectral analysis on the resulting data set was carried out by including {\tt\string gll\_iem\_v07} and the isotropic diffuse model {\tt\string iso\_P8R3\_SOURCE\_V2\_v1}.  A region of interest of $10^\circ$ centered at the source was used for analysis. To accommodate and model the photons coming from vicinity of the source of interest, the spectral parameters of sources within $5^\circ$ of the region of interest were allowed to vary.  Additionally, sources with a variability index $>25$ within $15^\circ$ were made to vary.  However, spectral parameters of  all the sources outside $5^\circ$ of the region of interest and with variability index $<25$  were fixed to their 4FGL catalog values.  The flux and spectrum of PKS~1424$-$418 were determined by fitting a log-parabola model, using a binned gtlike algorithm based on the NewMinuit optimizer. A test statistics (TS)$\,>9$ suggests a detection significance of more than $3\sigma$  ($\sqrt{\rm{TS}}\sim3$).  The  TS is defined as $\rm{TS}=-2\,ln(\mathcal{L}_0/\mathcal{L}_1)$ where $\mathcal{L}_0$ and $\mathcal{L}_1$ denote the likelihood value without and with the point source at the position of interest, respectively. 

\begin{figure*}  
\centering
\includegraphics [width=1.0\textwidth]{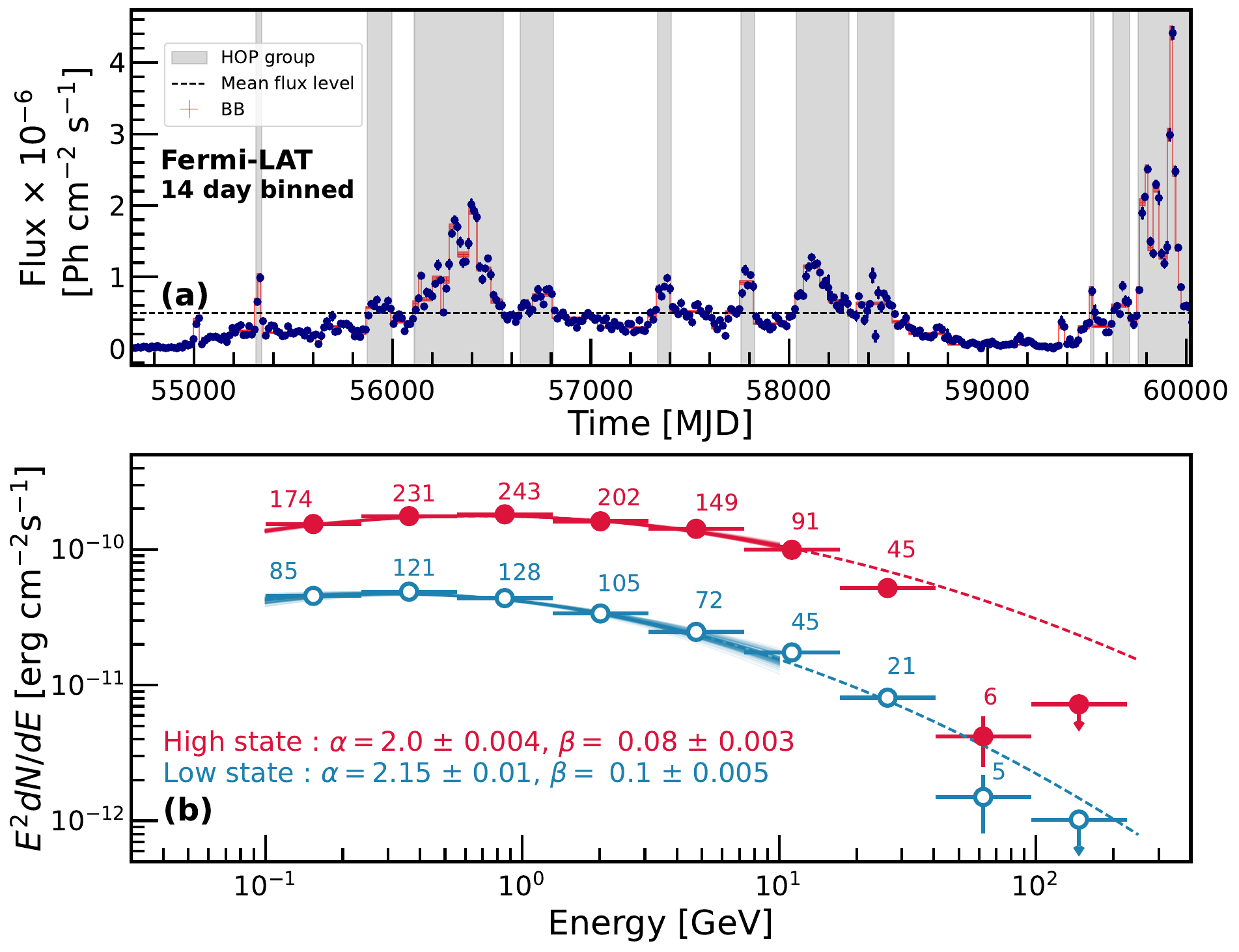}
\caption{ (a) 14 day binned light curve of $\approx$ 15 yr of Fermi-LAT observation of PKS~1424$-$418. The gray region represents high state and white represents low state (b) The combined spectrum of all ``flaring'' and ``quiescent'' epochs marked in panel (a). Intrinsic spectrum is modeled with a log-parabola model up to $10$GeV (solid line with envelope) and then extrapolated up to $\sim$300 GeV (dotted lines). The significance of detection ($\sqrt{\rm{TS}}$) for each energy bin is mentioned over the respective bins.  }
\label{fig:quiet_high_states}
\end{figure*}
\vspace{-2em}
\subsection{Identification of Bright flaring epochs in light curve (Bayesian Block and HOP Algorithm)} \label{subsec:HOP_BB_methods} Only periods of significant detection are further considered for analysis. In this work, a time bin in Fermi-LAT light curve is considered a detection if it has $\rm{TS}>9$ and if flux in one bin is greater than its uncertainty, i.e. $F_t>\sigma_t$. 

We represent the flux points and the associated uncertainties in step-function representation using Bayesian Block (BB) to detect and characterize variability structures localized in time \citep{Scargle_2013, 2023Agarwal}. Each point of change of the block in the step-function representation highlights the 3$\sigma$ variation from the previous block. 
To identify the flaring features in the light curve, the BB is fed into the HOP algorithm, which is based on a \textit{watershed} concept used in topological data analysis \citep{1998ApJ...498..137E}. 
This segregates the light-curve periods into multiple flaring groups (further identified as HOP flaring groups) and quiescent groups based on the emergence of BB above the mean flux level. 
This has been used in \citet{Meyer_2019} to zoom into periods of flaring epochs and identify signatures of compact emission regions in FSRQs. 
The flare identification code by \citet{Wagner:2021jn} is used to identify the different HOP groups. The above classification resulted in 11 HOP flaring groups and 11 quiescent periods as indicated in Fig. \ref{fig:quiet_high_states}(a). 

\subsection{Flux distribution} \label{subsec:flux_distribution_method}
The assessment of flux patterns for various flaring groups of PKS~1424$-$418 is conducted based on the techniques presented in the study by \citet{2021MNRAS.504.1427A} and \citet{2023Agarwal}. These flux profiles are then subjected to fitting processes using the following functions:  \\
1. Gaussian:
\vspace{-3mm}
\begin{equation}
\rm{G} (x;\mu_{\rm{G}},\sigma_{\rm{G}} )=\frac{N_{\rm{G}}}{\sigma_{\rm{G}}\sqrt{2\pi}}\exp\left[{-\frac{(x-\mu_{\rm{G}})^2}{2\sigma_{{\rm{G}}}^2}}\right]
\end{equation}
\\
2. Lognormal : 
\vspace{-3mm}
\begin{equation}
\rm{LN} (x;\mu_{{\rm{LN}}},\sigma_{{\rm{LN}}})=\frac{N_{{\rm{LN}}}}{x\sigma_{{\rm{LN}}}\sqrt{2\pi}}\exp\left[{-\frac{(\log(x)-\mu_{{\rm{LN}}})^2}{2\sigma_{{\rm{LN}}}^2}}\right]
\end{equation}

Here, $N_{i}$, $\mu_{i}$ and $\sigma_{i}$  are the normalization constant, mean, and standard deviation for the fitted profiles, respectively (i= G or LN indicating Gaussian or Lognormal profile).  The selection of the more favorable distribution fit is determined by considering the Akaike information criterion (AIC) values \citep{1974Akaike}. A lower AIC for a model means a better description of the data. 

\section{Result} \label{sec:result}
\subsection{Activity Periods}
A visual inspection of high-energy $\gamma$-ray ($0.1-300\,$GeV) light curve over $\sim$ 15 yr displays distinct periods of high activity and low activity. A significant highlight of this period was the extended flare in 2022, which manifested as a pronounced outburst observed in optical (ATOM; \citealt{2022ATel15552....1J}), radio (ATCA; \citealt{2022ATel15536....1K}), and $\gamma$-ray (Fermi-LAT and AGILE; \citealt{2022ATel15807....1L, 2022ATel15818....1V}) emissions. 
Flaring groups, denoted by gray patches in Fig. \ref{fig:quiet_high_states}(a), are interspersed with multiple quiescent groups, represented by white patches in Fig. \ref{fig:quiet_high_states}(a). These flaring and quiescent groups are identified using the HOP algorithm, as detailed in section \ref{subsec:HOP_BB_methods}.

We examine the collective high-energy $\gamma$-ray spectrum ($0.1\,-\,300\,$GeV) of all the photons during the ``flaring" and ``quiescent" periods. The culmination of all the flaring groups and quiescent groups are hereby referred to as the ``high state" and ``low state", respectively, in the paper. The $\gamma$-ray spectrum is modeled using a log-parabola model, parameterized as: 
\vspace{-0.3em}
\begin{equation}  \label{eq:3}
\frac{dN}{dE}=N_\circ \left( \frac{E}{E_b} \right)^{-(\alpha + \beta \log(E/E_b))}
\end{equation}

where $E_b$ was fixed to 4FGL catalog value of $677.45$\,MeV and $N_{o}$ is the Normalization. We also tested with the power-law model and found that $\gamma$-ray spectrum favors a log-parabola model over a power-law, with a TS value of 231 and 151 during high and low state, respectively. 

For PKS~1424$-$418, a redshift of $z=1.522$ implies a resulting attenuation due to optical-UV-near-IR extragalactic background light (EBL) radiation beyond critical energy, $E_{\rm{crit}} \approx 170 (1+z)^{-2.38}\, \rm{GeV}$ = $18.8\, \rm{GeV}$ \citep{Ackermann_science_2012}. 
\citet{Poutanen2010} found a significant break in bright blazars indicating $\gamma$-ray absorption via photon$-$photon pair production on helium II (He II) and hydrogen (H) recombination continuum photons. \citet{Stern_Poutanen_2014} observed a significant 20 GeV break in the source frame due to the H Lyman continuum (LyC) using revised Pass 7 Fermi-LAT response function and found that any break due to He II is less significant. Any spectral features in soft spectrum at energy $E_{\rm{soft}}$ interact with $\gamma$-ray photons at energy $E_{\rm{hard}}$ and are manifested as an observed attenuation beyond the threshold energy
\begin{equation} \label{eq:4}
E_{\rm{th}} \gtrsim \frac{(m_e c^2)^2}{E_{\rm{soft}} (1+z) (1-\cos\,\theta)} \simeq 10\, \left( \frac{10\, \rm{eV}}{E_{\rm{soft,Ly}\alpha}} \right) \rm{GeV}
\end{equation}

which is the minimum energy for absorption in a head-on collision ($\theta=180^{\circ}$). BLR photons may influence the high-energy spectrum only beyond $10\,$GeV (equation \ref{eq:4}). Thus, the spectrum below $\approx$ $10\,$GeV is a true estimate of the unabsorbed intrinsic spectrum of the source.
 
Combined high-energy $\gamma$-ray spectrum of high and low states up to 10 GeV are fitted with log-parabola model using equation \ref{eq:3} and have consistent values of $\beta$ parameters as in Fig. \ref{fig:quiet_high_states}(b). Since EBL attenuation and BLR influence is relevant only beyond 10 GeV, consistency in the $\beta$ parameter for low- and high-state spectrum hints at a similar influence of external seed photons up to 10 GeV. 
\vspace{-0.5em}
\subsection{ Fast variability}
\begin{figure}
\centering
\includegraphics [width=0.5\textwidth]{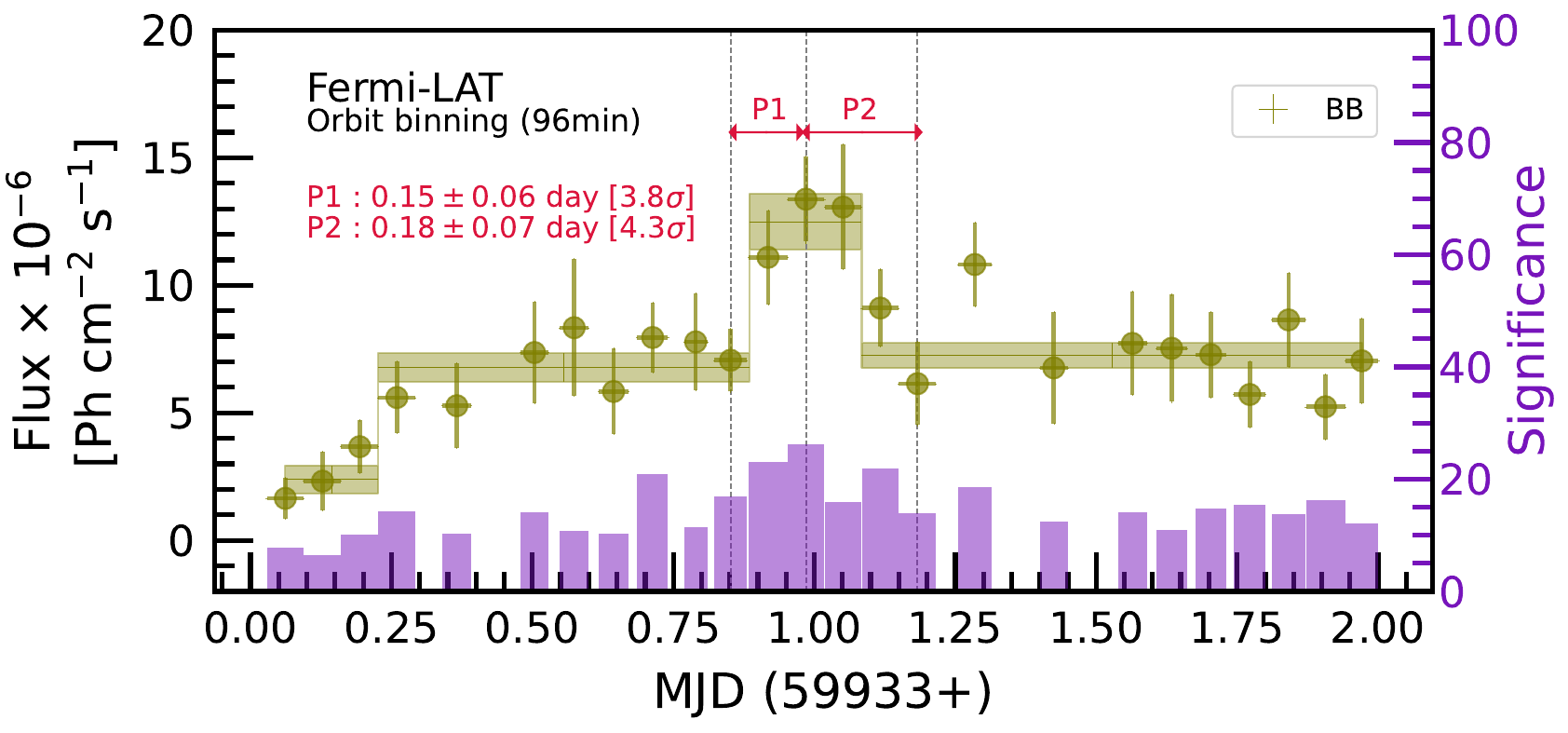}
\caption{Orbit-binned light curve of a period of fastest variability in Fermi-LAT of PKS~1424$-$418. Bayesian Blocks with a false alarm probability of 5\% indicating significant points of change are plotted on top. The periods of significant variability are marked with grey dotted lines. }
\label{fig:fast_variability}
\end{figure}

During the most intense flaring episodes in 2022, the flux of PKS~1424$-$418 surged to approximately 10 times the $\approx15$ yr flux average. On 2022 December 20, Fermi-LAT recorded a remarkably fast variability during the source's brightest flare. The orbit-binned light curve ($\leq\,$96 minutes) in Fig. \ref{fig:fast_variability} illustrates the variable flux, with its 3$\sigma$  variation represented by the corresponding BB. Recently, Fermi-LAT captured a significant (3.8$\,\sigma$) intraday variability of 0.15~$\pm$~0.06 days during the brightest flux state of the source (Fig. \ref{fig:fast_variability}).

The source is visibly variable during the brightest activity period between MJD 59760 and 59961, statistically evident from the pink noise behavior down to 6 hr (power-law index $\approx1.16\pm0.25$, details in Agarwal et al 2024, in preparation). With a black hole mass of $M_{\rm{BH}}=4.5\times 10^{9}\,M_{\odot}$, light-crossing timescales of $\approx$ 0.5 days is a measure of minimum expected variability in the jet frame \citep{2001spada}.

Between MJD 59758 and 60024, the flux doubled in $(t_{\rm{var}})_{\rm{obs}}=0.15\,\pm\,0.06$ days shown in Fig. \ref{fig:fast_variability}. We quantify the variability timescales associated with these points of change using $t_{\rm{var}}=(t_2 - t_1)\frac{\ln 2}{\ln(F_2/F_1)}$ \citep{2011A&A...530A..77F}, where $F_2$ and $F_1$ are the fluxes at time $t_2$ and $t_1$ respectively, and $t_{\rm{var}}$ is the flux doubling and halving timescales. Such short variability timescales place a tight constraint on the size and location of the emission region. Here, the minimum size of the emission region is of the same order as the size of the black hole. This is the fastest-ever recorded variability in the source to date. Previously, \citet{2021MNRAS.501.2504A} observed a variability of 3.6 days (4.74$\,\sigma$) during 2012 in the variable periods of MJD 56015$–$56020. The observed variability constrains the emission blob at a maximum radius of $r'_{\rm{emm}}=\,ct_{\rm{var}}\delta/(1+z)=1.5\times10^{15}(\delta/10)\,$ cm. The typical dissipation distance from supermassive black hole for a radiation region of size $r'_{\rm{emm}}$, is  $R_{\rm{diss}}=2c\Gamma^2_j t_{\rm{var}}=0.025\,(\delta/10)^2\,$pc, assuming the Doppler factor of the blob  $\delta=\Gamma_j$ \citep{2019Galax...7...28R}.  Assuming the emission region covers the entire jet cross section, the 0.025 pc emission region could be in the vicinity of seed photons from BLR or accretion disk. 

\begin{deluxetable}{ccccc}
\tablenum{1}
\tablecaption{Parameters of flux distribution for Low  and High state\label{tab:flux_distribution}}
\tablewidth{0pt}
\tablehead{
\colhead{State} & \colhead{Model} &  \colhead{Mean ($\mu_{i}$)} & \colhead{Sigma ($\sigma_{i}$)} & \colhead{AIC}}
\decimalcolnumbers
\startdata
        \multirow{2}{*}{High state} & Gaussian & $3.74\pm0.01$  & $2.00\pm0.01$ & $-55999.5$\\
          & Lognormal & $1.43\pm0.01$ & $0.57\pm0.01$ & $-65411.3$\\
         \multirow{2}{*}{Low state} & Gaussian  & $7.36\pm0.01$ & $4.16 \pm0.01$ & $-47024.6$ \\
         & Lognormal & $2.11\pm0.01$ & $0.51\pm0.01$ & $-35417.7$\\
\enddata
\tablecomments{(1) The state of the light curves ---high state or low state.
(2) The models tested on the obtained flux distribution.
(3) Mean of the fitted model.
(4) Standard deviation of the fitted model.
(5) AIC values for the fitted model.}
\end{deluxetable}
\vspace{-1em}
\subsection{Flux Distribution}
The evaluation of flux patterns across high and low state is performed using flux distribution as described in section \ref{subsec:flux_distribution_method}. The AIC value suggests high states prefer a log-normal distribution, contrasting with the Gaussian distribution for low states (see Table \ref{tab:flux_distribution}).
Lognormal fluctuation hints at a multiplicative process at play, typically known for accreting galactic sources like X-ray binaries. This provides hints of the influence of accretion disk on the jet \citep{2005MNRAS.359..345U}, or a minijet in the jet model due to the Pareto distribution \citep{2012A&A...548A.123B} possibly due to magnetic reconnection scenario at the edge of the BLR as studied in \citet{2023Agarwal}.

\subsection{EBL attenuation}
The unabsorbed intrinsic spectrum of the source is estimated from $0.1-10$ GeV and extrapolated further to higher energies up to 100 GeV as in Fig. \ref{fig:quiet_high_states}(b). We use reduced chi-square ($\chi^{2}$) to test if the intrinsic spectrum is a true representation of the observed spectrum up to 100 GeV as in \citet{10.1093/mnras/sty887}. For the high state, the intrinsic spectrum is rejected with a $p$-value $<$ $10^{-5}$. For low-state, the model is rejected with a $p$-value $\sim\,10^{-4}$. Notably, the main contribution for high $\chi^{2}$ comes from the high-energy end ($E>10$ GeV) of the spectrum (see Fig. \ref{fig:quiet_high_states}(b)).

The observed spectrum shows a contrasting deviation from the extrapolated intrinsic fit, with a  21.8$\sigma$ deviation for photons in the high state and a 2.8$\sigma$ deviation in the low state within the range of 40$-$95 GeV. This deviation is most likely caused by the absorption of $\gamma$-ray photons interacting with EBL photons alone or with EBL photons in combination with soft photons from the local jet environment (accretion disk, torus, or BLR) during their journey to the observer.

To account for EBL absorption, the extrapolated intrinsic spectrum is modified with an exponential term, $e^{-\tau(E,z)}_{\gamma,\gamma}$ , such that $\,F_{\rm{obs}}(E)\, =\, F_{\rm{int}}(E)\, \exp [-\tau_{\gamma,\gamma}(E,z)]$ \citep{2004Kneiske}. Here, $\tau_{\gamma,\gamma}(E,z) = b \times \tau^{\rm{model}}_{\gamma,\gamma}(E,z)$, where  $\tau^{\rm{model}}_{\gamma,\gamma}(E,z)$ is the predicted optical depth by various EBL models and $b$ represents the opacity scaling factor. 
Considering the level of EBL absorption as described by the model, the value of $b$ highlights either of two conditions: (1) $b=0$ highlights no EBL attenuation, (2) $b=1$ indicates correct model selection, hence the correct estimate of EBL absorption.

\begin{figure*}  
\centering
\includegraphics [width=1.0\textwidth]{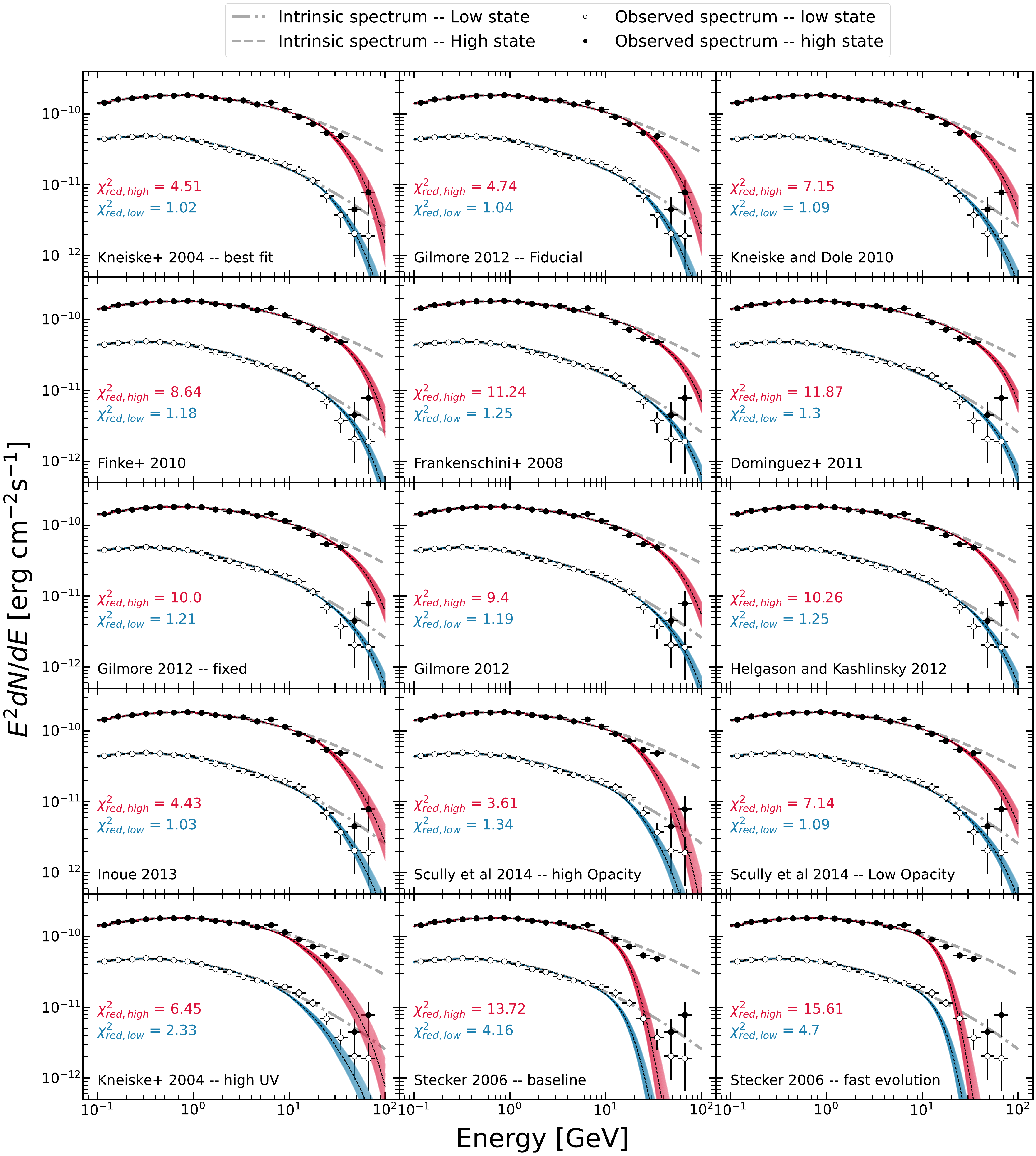}
\caption{ Observed $\gamma$-ray spectrum in 0.1$-$100 GeV range for the high state and low state. The true intrinsic spectrum in the energy range of 0.1$-$10 GeV extrapolated up to 100 GeV for the high and low state is shown in gray dashed and dotted-dashed lines, respectively. The solid red and blue envelope represents the intrinsic high-state and low-state spectrum, respectively, after EBL attenuation using opacity scaling factor (b) ranging from 0.75 to 1.25 across different models indicated in each panel. The mean value of the fit corresponding to $b$=1 is indicated by a black dashed line. The quoted reduced $\chi^{2}$ in each panel is for the best-fit corresponding to $b$=1.}
\label{fig:best_fit_model_kneiske}
\end{figure*}

We examined the observed high- and low-state spectrum with 15 EBL-absorbed spectra ($F_{obs}$) with $b$ value ranging from 0.75 - 1.25 to accommodate 25\% tolerance on the optical depths \citep{Ackermann_science_2012} predicted by models at different energies ( \citealt{2004Kneiske}, \citealt{2012Gilmore}, \citealt{2010Kneiske}, \citealt{2010Finke}, \citealt{2008Franceschini}, \citealt{2011Dominguez}, \citealt{2012Helgason}, \citealt{2013Inoue}, \citealt{2014Scully}, \citealt{2006Stecker}).
Among the 15 EBL-absorbed spectra derived from 15 models above, 12 models resulted in $\chi^{2}$ close to 1 in the low state, but follow a deviation with a significantly high $\chi^{2}$ in the high state as listed in Fig. \ref{fig:best_fit_model_kneiske}. Since EBL is uniform and isotropic on large scales, the expected levels of $\gamma$-ray absorption from a source should be flux-independent. Thus, the high and low states are expected to have similar levels of absorption from the photon$-$photon absorption in presence of EBL photons. However, none of the present EBL models can efficiently explain the absorption levels beyond 10 GeV for the high state using EBL absorption alone. A certain level of extra absorption is imprinted on the high-state spectrum reflecting an additional effect of intervening interacting photons. Conservatively, the observed high-state spectrum deviates by $\sim4.7\sigma$ beyond $10$ GeV from the EBL-absorbed spectra derived using the \citet{2014Scully} -- high-opacity model. Additionally, for the more widely used EBL model, by \citet{2008Franceschini} and \citet{2011Dominguez}, the absorption significance beyond E$>10\,$GeV are found to be $>5\sigma$.

\vspace{-1em}
\section{Discussion} \label{sec:discussion}
Spectral breaks in $\gamma$-ray spectrum are expected at different energies, owing to (a) internal absorption through photon$-$photon interactions with external seed photons, primarily from the BLR, accretion disk, and dusty torus, and (b) absorption of high-energy photons by the EBL in the optical, UV, and near-IR bands. Such multiple absorptions make the detection of high-energy photons in high-redshift objects extremely challenging due to poor photon statistics. We report a significant deviation of the stacked high-state and low-state spectrum from the fitted log-parabola model at $E>$ 10 GeV as evident in Fig. \ref{fig:quiet_high_states}(b). 

At energy beyond 10 GeV, two energy bins have greater than $3\sigma$ deviation in the high state (Fig. \ref{fig:quiet_high_states}(b)). During the low state of the source, the deviation is limited to 0.6$\sigma$ within $17-40$ GeV and $2.8\sigma$, within $40-95$ GeV. This deviation is significantly smaller than 4$\sigma$ and 21.8$\sigma$ deviation within $17-40$ GeV and $40-95$ GeV, respectively, during the high state of the source (Fig. \ref{fig:quiet_high_states}(b)). This deviation carries the imprint of absorbed high-energy photons during the high states of the source.
The imprint includes absorption from both EBL and external jet photons (i.e. BLR, torus, or accretion disk), contingent on the emission site's position during high states. The absence of the absorption feature in the low state indicates that external photons, apart from the EBL, could have a notable impact on the $\gamma$-ray spectrum during high states. Low-activity periods are typically associated with the outer parsec-scale regions of the jet or result from combined emissions along the entire jet length in the absence of a dominant emission zone. High-activity periods are primarily linked to emission originating from energetic particles within the inner jet at parsec scales from black hole \citep{2022ApJ...939...76E}. Additionally, high Compton dominance in the source ($q\,\sim30$; \citealt{2021MNRAS.501.2504A}) indicates that accelerated high-energy electrons in the jet scatter a fraction of soft photons, emitting $\gamma$-rays. This process necessitates proximity of high-energy electrons to dominant sites of the soft photons, such as the accretion disk, BLR, or dusty torus. 
The observed variability of 0.15 $\pm$ 0.06 days constrains the emission region's location to be at a minimum distance of $R_{\rm{diss}}>0.025$ pc from the central black hole.  In Agarwal et al. (2025, in preparation), a $E_{\rm{HE,max}}=65\,$GeV photon, alongside $\sim$ 200 photons with $E>10$ GeV, was detected with over 99\% probability of association with PKS~1424$-$418 during the 2022 flaring activity. For a flat BLR, $\gamma$-ray emission site is constrained at  $r_{\rm{min}}=r_{\rm{BLR}}/\tan\theta_{\rm{min}}\simeq\,0.45\,\rm{pc}$  (\citealt{2012MNRAS.425.2519N}) from a central super massive black hole where $r_{\rm{BLR}}=0.5\,$pc (using $L_{\rm{disk}}=2.5\times10^{47}\,\rm{erg\,s}^{-1}$ as in \citealt{2021MNRAS.501.2504A} and \citealt{ 2014A&A...569A..40B} and $r_{\rm{BLR}}=0.1\,\rm{pc}$$\times(L_{\rm{disk}}/10^{46})^{1/2}$ from \citealt{2012MNRAS.425.2519N}). The minimum collision angle $\theta_{\rm{min}}$ at $E_{\rm{th}}=E_{\rm{HE,max}}$ is given from equation \ref{eq:4} as,

\begin{equation}
\theta_{\rm{min}}=\arccos\left(1-\frac{2(m_e c^2)^2}{(1+z)E_{\rm{HE,max}} E_{\rm{soft,Ly}\alpha}}\right)\simeq47^{\circ}.
\end{equation}

At 0.45 pc, both the BLR and accretion disk could contribute as sources of external seed photons. However, if the BLR has a significant `tail' for $r>r_{\rm{BLR}}$, the high-energy emission is produced at least within the parsec scale.

Hence, the observed attenuation could be attributed to the ``local opacity" effect caused by the BLR and accretion disk photon. To explore this further, we investigate potential sources of change in $\gamma$-ray opacity in a local jet scenario.  
\vspace{-1em}
\subsection{Imprints of BLR }
Due to varying degrees of ionization, the BLR emits several strong line features. \citet{Poutanen2010} synthesize the most significant features in the BLR spectrum and its effect on high-energy photon propagation. The interaction of BLR line photons with jet photons could result in photon$-$photon absorption at energy threshold of pair production. The energy threshold for photon$-$photon pair production depends on collision angle as well energy of soft seed photon, such that, $E_{\rm{th}}\propto1/ E_{\rm{soft}}(1-\cos\theta)$. Thus, the role of angular distribution of external radiation near the emission zone becomes important, emphasizing the contribution of BLR geometry and position of emission site with respect to BLR for changing opacity of $\gamma$-ray radiation (\citealt{2012arXiv1209.2291T}, \citealt{ lei2014}, \citealt{2017MNRAS.464..152A}). 
For an emission region within BLR at $r=R_{\rm{BLR}}$, isotropically distributed BLR radiation around the emission region leads to increased optical depth. Toward the outer edge of the BLR, the optical depth decreases rapidly, due to a shift from head-on collisions ($\theta=180^{\circ}$) to less favorable angles of $\theta<90\degr$ \citep{2012arXiv1209.2291T}. This decrease is much more pronounced near the opacity threshold energy \citep{2017MNRAS.464..152A}. Typically, spectral breaks are imprinted on the high-energy spectrum by two emission lines, hydrogen Ly$\alpha$ (H Ly$\alpha$) and helium II Ly$\alpha$ (He Ly$\alpha$), produced near the central engine. \citet{2017MNRAS.464..152A} emphasized contributions from various emission lines at larger distances, dominated mostly by low-ionization lines. 

 At $r_{\rm{min}}=0.45\,$pc, toward the edge of the BLR, low-ionization lines become more influential. This is a result of large collision angles, in contrast to the small collision angles associated with high-ionization lines localized within the inner boundaries of the BLR. Absorption in the source frame, around $\sim10-30$ GeV, may result from increase in $\gamma$-ray opacity due to H Ly$\alpha$ and LyC (with $E_{\rm{BLR,Ly}\alpha}=10.2$ eV and $E_{\rm{BLR,LyC}}=13.6$ eV). While the resulting break could shift to higher energies, efficiency of photon$-$photon absorption may be reduced with decreased collision angles. The observed absorption around $100/(1+z)\,$-$\,140/(1+z)\,$GeV cannot be attributed to the dominant H Ly$\alpha$ and LyC in the BLR. However, several strong lines at lower energies, such as Balmer H$\alpha$ and H$\beta$ (with $E_{\rm{BLR,H}\alpha}=1.89$eV and $E_{\rm{BLR,H}\beta}=2.55\,$eV), may contribute to the noted absorption within this energy range \citep{2013APh....43..215S}. Interestingly, the position of emission site at 0.45 pc, at the outer edge of the BLR, should be dominated by the low-ionization Balmer lines resulting in absorption at $40-95$ GeV in observer's frame. Toward the edge of the flat BLR, head-on collisions are significantly reduced as collision angles are limited to $\theta<90\degr$. This should result in doubling the break energy \citep{Stern_Poutanen_2014}. However, if a few BLR photons are positioned along the jet axis for head-on collisions outside the $\gamma$-ray emitting region, the observed absorption at $40-95$ GeV due to $H\alpha$, $H\beta$ could be justified. We implemented the BLR cutoff on the EBL-absorbed high-state spectrum, employing the commonly used EBL model by \citet{2008Franceschini}. The fitting highlighted the impact of low-ionization lines, as indicated by a significant improvement in the $\chi^2$ from $\sim11.2$ to $\sim2.5$, with the inferred cutoff at 44 GeV. The reduced $\chi^{2}$ larger than 1 for the BLR cutoff model is possible due to the choice of a single exponential cutoff model, which could be too simplistic for the much more complex contribution from the BLR lines and continua.

\vspace{-3mm}
\subsection{Imprints of Accretion disk}
Absorption beyond $10$ GeV indicates soft photon energies causing attenuation are within $E_{\rm{soft}}=$ 1 to 10 eV. While EBL origin would imply consistent absorption across flux levels, higher absorption during the flaring state suggests a different origin for these optical photons, possibly from accretion disk at the jet's base. PKS~1424$-$418 displays a bright disk emission, dominant within the optical range \citep{2021MNRAS.501.2504A, 2014A&A...569A..40B}. 

A lognormal distribution of flux in high-energy $\gamma$-rays during high state, in contrast with Gaussian distribution during low state implies distinct emission characteristics in two states. The preferred lognormal distribution indicates the possible influence of disk photons on jet specifically during high states \citep{2019Galax...7...28R} or a minijet in jet model due to the magnetic reconnection scenario at the edge of the BLR \citep{2012A&A...548A.123B, 2023Agarwal}.

At $r_{\rm{min}}\,=\,0.45$ pc, the radiation field from the accretion disks (dominant up to $\approx 10^{17} $ cm $\sim 0.03 $ pc) has a minimal contribution to the opacity of $\gamma$-rays. The alignment of the soft disk photons with the direction of the $\gamma$-rays from the jet reduces their interaction rates significantly, making the imprint of accretion disk photons on the high-energy spectrum highly unlikely \citep{2017MNRAS.464..152A}. Additionally, for reasonable Thomson optical depths of the hot intercloud medium, the disk photons scattering off free electrons also do not significantly contribute to the photon$-$photon pair production optical depth. Furthermore, unabsorbed photons, postaccretion disk influence, undergo additional absorption from BLR photons scattering at larger angles, resulting in higher opacity.
Consequently, the observed $\gamma$-ray site is unlikely to be within the inner radius of the BLR. 
\vspace{-3mm}
\section{Summary}  \label{sec:summary}
In this Letter, we report significant absorption of $\gamma$-ray photons at energies exceeding 10 GeV, possibly due to photon$-$photon pair production on low-ionization BLR photons from the outer edge of the BLR. We detected this absorption signature in PKS~1424$-$418  with high significance during a high state.
Conversely, the absence of this absorption signature in the low state suggests the presence of an emission region located further away from the BLR. The absorption feature during the high state, in contrast with the absence of it during the low state, supports the interpretation that powerful emission events originating within or at the edge of the BLR evolve into fainter emission components outside of the BLR. This is further supported by the detection of fast variability during the high state consistent with a size of the emission site of 0.45 pc which is within the outer edge of the BLR. 

\begin{acknowledgements}
{\bf Acknowledgements}
We thank the anonymous referee for constructive criticism, which has helped improve the manuscript. This research work has made use of data, software and web tools obtained from NASA’s High Energy Astrophysics Science Archive Research Center (HEASARC) and Fermi gamma-ray telescope Support centre, a service of the Goddard Space Flight Center and the Smithsonian Astrophysical Observatory. BB acknowledges financial support from the Italian Ministry of University and Research (MUR) for the PRIN grant METE under contract no. 2020KB33TP.
\end{acknowledgements}

\vspace{-2em}
\bibliography{main}{}

\begin{thebibliography}{}
\expandafter\ifx\csname natexlab\endcsname\relax\def\natexlab#1{#1}\fi
\providecommand{\url}[1]{\href{#1}{#1}}
\providecommand{\dodoi}[1]{doi:~\href{http://doi.org/#1}{\nolinkurl{#1}}}
\providecommand{\doeprint}[1]{\href{http://ascl.net/#1}{\nolinkurl{http://ascl.net/#1}}}
\providecommand{\doarXiv}[1]{\href{https://arxiv.org/abs/#1}{\nolinkurl{https://arxiv.org/abs/#1}}}

\bibitem[{{Abeysekara} {et~al.}(2015){Abeysekara}, {Archambault}, {Archer}, {Aune}, {Barnacka}, {Benbow}, {Bird}, {Biteau}, {Buckley}, {Bugaev}, {Cardenzana}, {Cerruti}, {Chen}, {Christiansen}, {Ciupik}, {Connolly}, {Coppi}, {Cui}, {Dickinson}, {Dumm}, {Eisch}, {Errando}, {Falcone}, {Feng}, {Finley}, {Fleischhack}, {Flinders}, {Fortin}, {Fortson}, {Furniss}, {Gillanders}, {Griffin}, {Grube}, {Gyuk}, {H{\"u}tten}, {H{\r{a}}kansson}, {Hanna}, {Holder}, {Humensky}, {Johnson}, {Kaaret}, {Kar}, {Kelley-Hoskins}, {Khassen}, {Kieda}, {Krause}, {Krennrich}, {Kumar}, {Lang}, {Maier}, {McArthur}, {McCann}, {Meagher}, {Moriarty}, {Mukherjee}, {Nieto}, {O'Faol{\'a}in de Bhr{\'o}ithe}, {Ong}, {Otte}, {Park}, {Perkins}, {Petrashyk}, {Pohl}, {Popkow}, {Pueschel}, {Quinn}, {Ragan}, {Ratliff}, {Reynolds}, {Richards}, {Roache}, {Rousselle}, {Santander}, {Sembroski}, {Shahinyan}, {Smith}, {Staszak}, {Telezhinsky}, {Todd}, {Tucci}, {Tyler}, {Vassiliev}, {Vincent}, {Wakely}, {Weiner}, {Weinstein}, {Wilhelm}, {Williams}, {Zitzer},
  {VERITAS}, {Smith}, {SPOL}, {Holoien}, {Prieto}, {Kochanek}, {Stanek}, {Shappee}, {ASAS-SN}, {Hovatta}, {Max-Moerbeck}, {Pearson}, {Reeves}, {Richards}, {Readhead}, {OVRO}, {Madejski}, {NuSTAR}, {Djorgovski}, {Drake}, {Graham}, {Mahabal}, \& {CRTS}}]{2015ApJ...815L..22A}
{Abeysekara}, A.~U., {Archambault}, S., {Archer}, A., {et~al.} 2015, \apjl, 815, L22, \dodoi{10.1088/2041-8205/815/2/L22}

\bibitem[{{Abhir} {et~al.}(2021){Abhir}, {Joseph}, {Pgm}, \& {Bose}}]{2021MNRAS.501.2504A}
{Abhir}, J., {Joseph}, J., {Pgm}, S.~R., \& {Bose}, D. 2021, \mnras, 501, 2504, \dodoi{10.1093/mnras/staa3639}

\bibitem[{{Abolmasov} \& {Poutanen}(2017)}]{2017MNRAS.464..152A}
{Abolmasov}, P., \& {Poutanen}, J. 2017, \mnras, 464, 152, \dodoi{10.1093/mnras/stw2326}

\bibitem[{{Acciari} {et~al.}(2021){Acciari}, {Ansoldi}, {Antonelli}, {Asano}, {Babi{\'c}}, {Banerjee}, {Baquero}, {de Almeida}, {Barrio}, {Becerra Gonz{\'a}lez}, {Bednarek}, {Bellizzi}, {Bernardini}, {Bernardos}, {Berti}, {Besenrieder}, {Bhattacharyya}, {Bigongiari}, {Blanch}, {Bonnoli}, {Bo{\v{s}}njak}, {Busetto}, {Carosi}, {Ceribella}, {Cerruti}, {Chai}, {Chilingarian}, {Cikota}, {Colak}, {Colombo}, {Contreras}, {Cortina}, {Covino}, {D'Amico}, {D'Elia}, {Da Vela}, {Dazzi}, {De Angelis}, {De Lotto}, {Delfino}, {Delgado}, {Delgado Mendez}, {Depaoli}, {Di Girolamo}, {Di Pierro}, {Di Venere}, {Do Souto Espi{\~n}eira}, {Dominis Prester}, {Donini}, {Doro}, {Fallah Ramazani}, {Fattorini}, {Ferrara}, {Foffano}, {Fonseca}, {Font}, {Fruck}, {Fukami}, {Garc{\'\i}a L{\'o}pez}, {Garczarczyk}, {Gasparyan}, {Gaug}, {Giglietto}, {Giordano}, {Gliwny}, {Godinovi{\'c}}, {Green}, {Green}, {Hadasch}, {Hahn}, {Heckmann}, {Herrera}, {Hoang}, {Hrupec}, {H{\"u}tten}, {Inada}, {Inoue}, {Ishio}, {Iwamura}, {Jormanainen}, {Jouvin},
  {Kajiwara}, {Karjalainen}, {Kerszberg}, {Kobayashi}, {Kubo}, {Kushida}, {Lamastra}, {Lelas}, {Leone}, {Lindfors}, {Lombardi}, {Longo}, {L{\'o}pez}, {L{\'o}pez-Coto}, {L{\'o}pez-Oramas}, {Loporchio}, {Machado de Oliveira Fraga}, {Maggio}, {Majumdar}, {Makariev}, {Mallamaci}, {Maneva}, {Manganaro}, {Maraschi}, {Mariotti}, {Mart{\'\i}nez}, {Mazin}, {Mender}, {Mi{\'c}anovi{\'c}}, {Miceli}, {Miener}, {Minev}, {Miranda}, {Mirzoyan}, {Molina}, {Moralejo}, {Morcuende}, {Moreno}, {Moretti}, {Munar-Adrover}, {Neustroev}, {Nigro}, {Nilsson}, {Ninci}, {Nishijima}, {Noda}, {Nozaki}, {Ohtani}, {Oka}, {Otero-Santos}, {Palatiello}, {Paneque}, {Paoletti}, {Paredes}, {Pavleti{\'c}}, {Pe{\~n}il}, {Perennes}, {Persic}, {Prada Moroni}, {Prandini}, {Priyadarshi}, {Puljak}, {Rhode}, {Rib{\'o}}, {Rico}, {Righi}, {Rugliancich}, {Saha}, {Sahakyan}, {Saito}, {Sakurai}, {Satalecka}, {Schleicher}, {Schmidt}, {Schweizer}, {Sitarek}, {{\v{S}}nidari{\'c}}, {Sobczynska}, {Spolon}, {Stamerra}, {Strom}, {Strzys}, {Suda}, {Suri{\'c}},
  {Takahashi}, {Tavecchio}, {Temnikov}, {Terzi{\'c}}, {Teshima}, {Torres-Alb{\`a}}, {Tosti}, {Truzzi}, {van Scherpenberg}, {Vanzo}, {Vazquez Acosta}, {Ventura}, {Verguilov}, {Vigorito}, {Vitale}, {Vovk}, {Will}, {Zari{\'c}}, {Arbet-Engels}, {Baack}, {Balbo}, {Beck}, {Biederbeck}, {Biland}, {Bretz}, {Bruegge}, {Buss}, {Dorner}, {Elsaesser}, {Hildebrand}, {Iotov}, {Klinger}, {Mannheim}, {Neise}, {Neronov}, {Noethe}, {Paravac}, {Rhode}, {Schleicher}, {Sliusar}, {Theissen}, {Walter}, {Valverde}, {Horan}, {Giroletti}, {Perri}, {Verrecchia}, {Leto}, {Sadun}, {Moody}, {Joner}, {Marscher}, {Jorstad}, {L{\"a}hteenm{\"a}ki}, {Tornikoski}, {Ramakrishnan}, {J{\"a}rvel{\"a}}, {Vera}, {Righini}, \& {Lien}}]{2021MNRAS.504.1427A}
{Acciari}, V.~A., {Ansoldi}, S., {Antonelli}, L.~A., {et~al.} 2021, \mnras, 504, 1427, \dodoi{10.1093/mnras/staa3727}

\bibitem[{Ackermann {et~al.}(2012)Ackermann, Ajello, Allafort, Schady, Baldini, Ballet, Barbiellini, Bastieri, Bellazzini, Blandford, Bloom, Borgland, Bottacini, Bouvier, Bregeon, Brigida, Bruel, Buehler, Buson, Caliandro, Cameron, Caraveo, Cavazzuti, Cecchi, Charles, Chaves, Chekhtman, Cheung, Chiang, Chiaro, Ciprini, Claus, Cohen-Tanugi, Conrad, Cutini, D’Ammando, de~Palma, Dermer, Digel, do~Couto~e Silva, Domínguez, Drell, Drlica-Wagner, Favuzzi, Fegan, Focke, Franckowiak, Fukazawa, Funk, Fusco, Gargano, Gasparrini, Gehrels, Germani, Giglietto, Giordano, Giroletti, Glanzman, Godfrey, Grenier, Grove, Guiriec, Gustafsson, Hadasch, Hayashida, Hays, Jackson, Jogler, Kataoka, Knödlseder, Kuss, Lande, Larsson, Latronico, Longo, Loparco, Lovellette, Lubrano, Mazziotta, McEnery, Mehault, Michelson, Mizuno, Monte, Monzani, Morselli, Moskalenko, Murgia, Tramacere, Nuss, Greiner, Ohno, Ohsugi, Omodei, Orienti, Orlando, Ormes, Paneque, Perkins, Pesce-Rollins, Piron, Pivato, Porter, Rainò, Rando, Razzano,
  Razzaque, Reimer, Reimer, Reyes, Ritz, Rau, Romoli, Roth, Sánchez-Conde, Sanchez, Scargle, Sgrò, Siskind, Spandre, Spinelli, Łukasz Stawarz, Suson, Takahashi, Tanaka, Thayer, Thompson, Tibaldo, Tinivella, Torres, Tosti, Troja, Usher, Vandenbroucke, Vasileiou, Vianello, Vitale, Waite, Winer, Wood, \& Wood}]{Ackermann_science_2012}
Ackermann, M., Ajello, M., Allafort, A., {et~al.} 2012, Science, 338, 1190, \dodoi{10.1126/science.1227160}

\bibitem[{{Agarwal} {et~al.}(2023){Agarwal}, {Banerjee}, {Shukla}, {Roy}, {Acharya}, {Vaidya}, {Chitnis}, {Wagner}, {Mannheim}, \& {Branchesi}}]{2023Agarwal}
{Agarwal}, S., {Banerjee}, B., {Shukla}, A., {et~al.} 2023, MNRASL, 521, L53, \dodoi{10.1093/mnrasl/slad023}

\bibitem[{Akaike(1974)}]{1974Akaike}
Akaike, H. 1974, IEEE Transactions on Automatic Control, 19, 716, \dodoi{10.1109/TAC.1974.1100705}

\bibitem[{{Aleksi{\'c}} {et~al.}(2011{\natexlab{a}}){Aleksi{\'c}}, {Antonelli}, {Antoranz}, {Backes}, {Barrio}, {Bastieri}, {Becerra Gonz{\'a}lez}, {Bednarek}, {Berdyugin}, {Berger}, {Bernardini}, {Biland}, {Blanch}, {Bock}, {Boller}, {Bonnoli}, {Borla Tridon}, {Braun}, {Bretz}, {Ca{\~n}ellas}, {Carmona}, {Carosi}, {Colin}, {Colombo}, {Contreras}, {Cortina}, {Cossio}, {Covino}, {Dazzi}, {de Angelis}, {de Cea Del Pozo}, {de Lotto}, {Delgado Mendez}, {Diago Ortega}, {Doert}, {Dom{\'\i}nguez}, {Dominis Prester}, {Dorner}, {Doro}, {Elsaesser}, {Ferenc}, {Fonseca}, {Font}, {Fruck}, {Garc{\'\i}a L{\'o}pez}, {Garczarczyk}, {Garrido}, {Giavitto}, {Godinovi{\'c}}, {Hadasch}, {H{\"a}fner}, {Herrero}, {Hildebrand}, {Hose}, {Hrupec}, {Huber}, {Jogler}, {Klepser}, {Kr{\"a}henb{\"u}hl}, {Krause}, {La Barbera}, {Lelas}, {Leonardo}, {Lindfors}, {Lombardi}, {L{\'o}pez}, {Lorenz}, {Majumdar}, {Makariev}, {Maneva}, {Mankuzhiyil}, {Mannheim}, {Maraschi}, {Mariotti}, {Mart{\'\i}nez}, {Mazin}, {Meucci}, {Miranda}, {Mirzoyan},
  {Miyamoto}, {Mold{\'o}n}, {Moralejo}, {Nieto}, {Nilsson}, {Orito}, {Oya}, {Paoletti}, {Pardo}, {Paredes}, {Partini}, {Pasanen}, {Pauss}, {Perez-Torres}, {Persic}, {Peruzzo}, {Pilia}, {Pochon}, {Prada}, {Prada Moroni}, {Prandini}, {Puljak}, {Reichardt}, {Reinthal}, {Rhode}, {Rib{\'o}}, {Rico}, {R{\"u}gamer}, {R{\"u}ger}, {Saggion}, {Saito}, {Saito}, {Salvati}, {Satalecka}, {Scalzotto}, {Scapin}, {Schultz}, {Schweizer}, {Shayduk}, {Shore}, {Sillanp{\"a}{\"a}}, {Sitarek}, {Sobczynska}, {Spanier}, {Spiro}, {Stamerra}, {Steinke}, {Storz}, {Strah}, {Suri{\'c}}, {Takalo}, {Tavecchio}, {Temnikov}, {Terzi{\'c}}, {Tescaro}, {Teshima}, {Thom}, {Tibolla}, {Torres}, {Treves}, {Vankov}, {Vogler}, {Wagner}, {Weitzel}, {Zabalza}, {Zandanel}, \& {Zanin}}]{2011A&A...530A...4A}
{Aleksi{\'c}}, J., {Antonelli}, L.~A., {Antoranz}, P., {et~al.} 2011{\natexlab{a}}, \aap, 530, A4, \dodoi{10.1051/0004-6361/201116497}

\bibitem[{{Aleksi{\'c}} {et~al.}(2011{\natexlab{b}}){Aleksi{\'c}}, {Antonelli}, {Antoranz}, {Backes}, {Barrio}, {Bastieri}, {Becerra Gonz{\'a}lez}, {Bednarek}, {Berdyugin}, {Berger}, {Bernardini}, {Biland}, {Blanch}, {Bock}, {Boller}, {Bonnoli}, {Borla Tridon}, {Braun}, {Bretz}, {Ca{\~n}ellas}, {Carmona}, {Carosi}, {Colin}, {Colombo}, {Contreras}, {Cortina}, {Cossio}, {Covino}, {Dazzi}, {De Angelis}, {De Cea del Pozo}, {De Lotto}, {Delgado Mendez}, {Diago Ortega}, {Doert}, {Dom{\'\i}nguez}, {Dominis Prester}, {Dorner}, {Doro}, {Elsaesser}, {Ferenc}, {Fonseca}, {Font}, {Fruck}, {Garc{\'\i}a L{\'o}pez}, {Garczarczyk}, {Garrido}, {Giavitto}, {Godinovi{\'c}}, {Hadasch}, {H{\"a}fner}, {Herrero}, {Hildebrand}, {H{\"o}hne-M{\"o}nch}, {Hose}, {Hrupec}, {Huber}, {Jogler}, {Klepser}, {Kr{\"a}henb{\"u}hl}, {Krause}, {La Barbera}, {Lelas}, {Leonardo}, {Lindfors}, {Lombardi}, {L{\'o}pez}, {Lorenz}, {Makariev}, {Maneva}, {Mankuzhiyil}, {Mannheim}, {Maraschi}, {Mariotti}, {Mart{\'\i}nez}, {Mazin}, {Meucci}, {Miranda},
  {Mirzoyan}, {Miyamoto}, {Mold{\'o}n}, {Moralejo}, {Nieto}, {Nilsson}, {Orito}, {Oya}, {Paneque}, {Paoletti}, {Pardo}, {Paredes}, {Partini}, {Pasanen}, {Pauss}, {Perez-Torres}, {Persic}, {Peruzzo}, {Pilia}, {Pochon}, {Prada}, {Prada Moroni}, {Prandini}, {Puljak}, {Reichardt}, {Reinthal}, {Rhode}, {Rib{\'o}}, {Rico}, {R{\"u}gamer}, {Saggion}, {Saito}, {Saito}, {Salvati}, {Satalecka}, {Scalzotto}, {Scapin}, {Schultz}, {Schweizer}, {Shayduk}, {Shore}, {Sillanp{\"a}{\"a}}, {Sitarek}, {Sobczynska}, {Spanier}, {Spiro}, {Stamerra}, {Steinke}, {Storz}, {Strah}, {Suri{\'c}}, {Takalo}, {Tavecchio}, {Temnikov}, {Terzi{\'c}}, {Tescaro}, {Teshima}, {Thom}, {Tibolla}, {Torres}, {Treves}, {Vankov}, {Vogler}, {Wagner}, {Weitzel}, {Zabalza}, {Zandanel}, {Zanin}, {MAGIC Collaboration}, {Tanaka}, {Wood}, \& {Buson}}]{2011ApJ...730L...8A}
---. 2011{\natexlab{b}}, \apjl, 730, L8, \dodoi{10.1088/2041-8205/730/1/L8}

\bibitem[{{Atwood} {et~al.}(2009){Atwood}, {Abdo}, {Ackermann}, {Althouse}, {Anderson}, {Axelsson}, {Baldini}, {Ballet}, {Band}, {Barbiellini}, {Bartelt}, {Bastieri}, {Baughman}, {Bechtol}, {B{\'e}d{\'e}r{\`e}de}, {Bellardi}, {Bellazzini}, {Berenji}, {Bignami}, {Bisello}, {Bissaldi}, {Blandford}, {Bloom}, {Bogart}, {Bonamente}, {Bonnell}, {Borgland}, {Bouvier}, {Bregeon}, {Brez}, {Brigida}, {Bruel}, {Burnett}, {Busetto}, {Caliandro}, {Cameron}, {Caraveo}, {Carius}, {Carlson}, {Casandjian}, {Cavazzuti}, {Ceccanti}, {Cecchi}, {Charles}, {Chekhtman}, {Cheung}, {Chiang}, {Chipaux}, {Cillis}, {Ciprini}, {Claus}, {Cohen-Tanugi}, {Condamoor}, {Conrad}, {Corbet}, {Corucci}, {Costamante}, {Cutini}, {Davis}, {Decotigny}, {DeKlotz}, {Dermer}, {de Angelis}, {Digel}, {do Couto e Silva}, {Drell}, {Dubois}, {Dumora}, {Edmonds}, {Fabiani}, {Farnier}, {Favuzzi}, {Flath}, {Fleury}, {Focke}, {Funk}, {Fusco}, {Gargano}, {Gasparrini}, {Gehrels}, {Gentit}, {Germani}, {Giebels}, {Giglietto}, {Giommi}, {Giordano}, {Glanzman},
  {Godfrey}, {Grenier}, {Grondin}, {Grove}, {Guillemot}, {Guiriec}, {Haller}, {Harding}, {Hart}, {Hays}, {Healey}, {Hirayama}, {Hjalmarsdotter}, {Horn}, {Hughes}, {J{\'o}hannesson}, {Johansson}, {Johnson}, {Johnson}, {Johnson}, {Johnson}, {Kamae}, {Katagiri}, {Kataoka}, {Kavelaars}, {Kawai}, {Kelly}, {Kerr}, {Klamra}, {Kn{\"o}dlseder}, {Kocian}, {Komin}, {Kuehn}, {Kuss}, {Landriu}, {Latronico}, {Lee}, {Lee}, {Lemoine-Goumard}, {Lionetto}, {Longo}, {Loparco}, {Lott}, {Lovellette}, {Lubrano}, {Madejski}, {Makeev}, {Marangelli}, {Massai}, {Mazziotta}, {McEnery}, {Menon}, {Meurer}, {Michelson}, {Minuti}, {Mirizzi}, {Mitthumsiri}, {Mizuno}, {Moiseev}, {Monte}, {Monzani}, {Moretti}, {Morselli}, {Moskalenko}, {Murgia}, {Nakamori}, {Nishino}, {Nolan}, {Norris}, {Nuss}, {Ohno}, {Ohsugi}, {Omodei}, {Orlando}, {Ormes}, {Paccagnella}, {Paneque}, {Panetta}, {Parent}, {Pearce}, {Pepe}, {Perazzo}, {Pesce-Rollins}, {Picozza}, {Pieri}, {Pinchera}, {Piron}, {Porter}, {Poupard}, {Rain{\`o}}, {Rando}, {Rapposelli}, {Razzano},
  {Reimer}, {Reimer}, {Reposeur}, {Reyes}, {Ritz}, {Rochester}, {Rodriguez}, {Romani}, {Roth}, {Russell}, {Ryde}, {Sabatini}, {Sadrozinski}, {Sanchez}, {Sander}, {Sapozhnikov}, {Parkinson}, {Scargle}, {Schalk}, {Scolieri}, {Sgr{\`o}}, {Share}, {Shaw}, {Shimokawabe}, {Shrader}, {Sierpowska-Bartosik}, {Siskind}, {Smith}, {Smith}, {Spandre}, {Spinelli}, {Starck}, {Stephens}, {Strickman}, {Strong}, {Suson}, {Tajima}, {Takahashi}, {Takahashi}, {Tanaka}, {Tenze}, {Tether}, {Thayer}, {Thayer}, {Thompson}, {Tibaldo}, {Tibolla}, {Torres}, {Tosti}, {Tramacere}, {Turri}, {Usher}, {Vilchez}, {Vitale}, {Wang}, {Watters}, {Winer}, {Wood}, {Ylinen}, \& {Ziegler}}]{2009ApJ...697.1071A}
{Atwood}, W.~B., {Abdo}, A.~A., {Ackermann}, M., {et~al.} 2009, \apj, 697, 1071, \dodoi{10.1088/0004-637X/697/2/1071}

\bibitem[{{Biteau} \& {Giebels}(2012)}]{2012A&A...548A.123B}
{Biteau}, J., \& {Giebels}, B. 2012, \aap, 548, A123, \dodoi{10.1051/0004-6361/201220056}

\bibitem[{{Buson} {et~al.}(2014){Buson}, {Longo}, {Larsson}, {Cutini}, {Finke}, {Ciprini}, {Ojha}, {D'Ammando}, {Donato}, {Thompson}, {Desiante}, {Bastieri}, {Wagner}, {Hauser}, {Fuhrmann}, {Dutka}, {M{\"u}ller}, {Kadler}, {Angelakis}, {Zensus}, {Stevens}, {Blanchard}, {Edwards}, {Lovell}, {Gurwell}, {Wehrle}, \& {Zook}}]{2014A&A...569A..40B}
{Buson}, S., {Longo}, F., {Larsson}, S., {et~al.} 2014, \aap, 569, A40, \dodoi{10.1051/0004-6361/201423367}

\bibitem[{Costamante {et~al.}(2018)Costamante, Cutini, Tosti, Antolini, \& Tramacere}]{10.1093/mnras/sty887}
Costamante, L., Cutini, S., Tosti, G., Antolini, E., \& Tramacere, A. 2018, MNRAS, 477, 4749, \dodoi{10.1093/mnras/sty887}

\bibitem[{{Dom{\'\i}nguez} {et~al.}(2011){Dom{\'\i}nguez}, {Primack}, {Rosario}, {Prada}, {Gilmore}, {Faber}, {Koo}, {Somerville}, {P{\'e}rez-Torres}, {P{\'e}rez-Gonz{\'a}lez}, {Huang}, {Davis}, {Guhathakurta}, {Barmby}, {Conselice}, {Lozano}, {Newman}, \& {Cooper}}]{2011Dominguez}
{Dom{\'\i}nguez}, A., {Primack}, J.~R., {Rosario}, D.~J., {et~al.} 2011, \mnras, 410, 2556, \dodoi{10.1111/j.1365-2966.2010.17631.x}

\bibitem[{Donea \& Protheroe(2003)}]{10.1143/PTPS.151.186}
Donea, A.~C., \& Protheroe, R.~J. 2003, Progress of Theoretical Physics Supplement, 151, 186, \dodoi{10.1143/PTPS.151.186}

\bibitem[{{Eisenstein} \& {Hut}(1998)}]{1998ApJ...498..137E}
{Eisenstein}, D.~J., \& {Hut}, P. 1998, \apj, 498, 137, \dodoi{10.1086/305535}

\bibitem[{{Ezhikode} {et~al.}(2022){Ezhikode}, {Shukla}, {Dewangan}, {Pawar}, {Agarwal}, {Mathew}, \& {Krishna}}]{2022ApJ...939...76E}
{Ezhikode}, S.~H., {Shukla}, A., {Dewangan}, G.~C., {et~al.} 2022, \apj, 939, 76, \dodoi{10.3847/1538-4357/ac9627}

\bibitem[{{Finke} {et~al.}(2010){Finke}, {Razzaque}, \& {Dermer}}]{2010Finke}
{Finke}, J.~D., {Razzaque}, S., \& {Dermer}, C.~D. 2010, \apj, 712, 238, \dodoi{10.1088/0004-637X/712/1/238}

\bibitem[{{Foschini} {et~al.}(2011){Foschini}, {Ghisellini}, {Tavecchio}, {Bonnoli}, \& {Stamerra}}]{2011A&A...530A..77F}
{Foschini}, L., {Ghisellini}, G., {Tavecchio}, F., {Bonnoli}, G., \& {Stamerra}, A. 2011, \aap, 530, A77, \dodoi{10.1051/0004-6361/201117064}

\bibitem[{{Franceschini} {et~al.}(2008){Franceschini}, {Rodighiero}, \& {Vaccari}}]{2008Franceschini}
{Franceschini}, A., {Rodighiero}, G., \& {Vaccari}, M. 2008, \aap, 487, 837, \dodoi{10.1051/0004-6361:200809691}

\bibitem[{{Ghisellini} \& {Tavecchio}(2009)}]{2009MNRAS.397..985G}
{Ghisellini}, G., \& {Tavecchio}, F. 2009, \mnras, 397, 985, \dodoi{10.1111/j.1365-2966.2009.15007.x}

\bibitem[{{Ghisellini} {et~al.}(2011){Ghisellini}, {Tagliaferri}, {Foschini}, {Ghirlanda}, {Tavecchio}, {Della Ceca}, {Haardt}, {Volonteri}, \& {Gehrels}}]{2011MNRAS.411..901G}
{Ghisellini}, G., {Tagliaferri}, G., {Foschini}, L., {et~al.} 2011, \mnras, 411, 901, \dodoi{10.1111/j.1365-2966.2010.17723.x}

\bibitem[{{Gilmore} {et~al.}(2012){Gilmore}, {Somerville}, {Primack}, \& {Dom{\'\i}nguez}}]{2012Gilmore}
{Gilmore}, R.~C., {Somerville}, R.~S., {Primack}, J.~R., \& {Dom{\'\i}nguez}, A. 2012, \mnras, 422, 3189, \dodoi{10.1111/j.1365-2966.2012.20841.x}

\bibitem[{{H.~E.~S.~S. Collaboration} {et~al.}(2013){H.~E.~S.~S. Collaboration}, {Abramowski}, {Acero}, {Aharonian}, {Akhperjanian}, {Anton}, {Balenderan}, {Balzer}, {Barnacka}, {Becherini}, {Becker Tjus}, {Behera}, {Bernl{\"o}hr}, {Birsin}, {Biteau}, {Bochow}, {Boisson}, {Bolmont}, {Bordas}, {Brucker}, {Brun}, {Brun}, {Bulik}, {Carrigan}, {Casanova}, {Cerruti}, {Chadwick}, {Chaves}, {Cheesebrough}, {Colafrancesco}, {Cologna}, {Conrad}, {Couturier}, {Dalton}, {Daniel}, {Davids}, {Degrange}, {Deil}, {deWilt}, {Dickinson}, {Djannati-Ata{\"\i}}, {Domainko}, {O'C. Drury}, {Dubus}, {Dutson}, {Dyks}, {Dyrda}, {Egberts}, {Eger}, {Espigat}, {Fallon}, {Farnier}, {Fegan}, {Feinstein}, {Fernandes}, {Fernandez}, {Fiasson}, {Fontaine}, {F{\"o}rster}, {F{\"u}{\ss}ling}, {Gajdus}, {Gallant}, {Garrigoux}, {Gast}, {Giebels}, {Glicenstein}, {Gl{\"u}ck}, {G{\"o}ring}, {Grondin}, {Grudzi{\'n}ska}, {H{\"a}ffner}, {Hague}, {Hahn}, {Hampf}, {Harris}, {Hauser}, {Heinz}, {Heinzelmann}, {Henri}, {Hermann}, {Hillert}, {Hinton},
  {Hofmann}, {Hofverberg}, {Holler}, {Horns}, {Jacholkowska}, {Jahn}, {Jamrozy}, {Jung}, {Kastendieck}, {Katarzy{\'n}ski}, {Katz}, {Kaufmann}, {Kh{\'e}lifi}, {Klepser}, {Klochkov}, {Klu{\'z}niak}, {Kneiske}, {Kolitzus}, {Komin}, {Kosack}, {Kossakowski}, {Krayzel}, {Kr{\"u}ger}, {Laffon}, {Lamanna}, {Lefaucheur}, {Lemoine-Goumard}, {Lenain}, {Lennarz}, {Lohse}, {Lopatin}, {Lu}, {Marandon}, {Marcowith}, {Masbou}, {Maurin}, {Maxted}, {Mayer}, {McComb}, {Medina}, {M{\'e}hault}, {Menzler}, {Moderski}, {Mohamed}, {Moulin}, {Naumann}, {Naumann-Godo}, {de Naurois}, {Nedbal}, {Nguyen}, {Niemiec}, {Nolan}, {Ohm}, {de O{\~n}a Wilhelmi}, {Opitz}, {Ostrowski}, {Oya}, {Panter}, {Parsons}, {Paz Arribas}, {Pekeur}, {Pelletier}, {Perez}, {Petrucci}, {Peyaud}, {Pita}, {P{\"u}hlhofer}, {Punch}, {Quirrenbach}, {Raab}, {Raue}, {Reimer}, {Reimer}, {Renaud}, {de los Reyes}, {Rieger}, {Ripken}, {Rob}, {Rosier-Lees}, {Rowell}, {Rudak}, {Rulten}, {Sahakian}, {Sanchez}, {Santangelo}, {Schlickeiser}, {Schulz}, {Schwanke}, {Schwarzburg},
  {Schwemmer}, {Sheidaei}, {Skilton}, {Sol}, {Spengler}, {Stawarz}, {Steenkamp}, {Stegmann}, {Stinzing}, {Stycz}, {Sushch}, {Szostek}, {Tavernet}, {Terrier}, {Tluczykont}, {Trichard}, {Valerius}, {van Eldik}, {Vasileiadis}, {Venter}, {Viana}, {Vincent}, {V{\"o}lk}, {Volpe}, {Vorobiov}, {Vorster}, {Wagner}, {Ward}, {White}, {Wierzcholska}, {Wouters}, {Zacharias}, {Zajczyk}, {Zdziarski}, {Zech}, \& {Zechlin}}]{2013A&A...554A.107H}
{H.~E.~S.~S. Collaboration}, {Abramowski}, A., {Acero}, F., {et~al.} 2013, \aap, 554, A107, \dodoi{10.1051/0004-6361/201321135}

\bibitem[{{Helgason} \& {Kashlinsky}(2012)}]{2012Helgason}
{Helgason}, K., \& {Kashlinsky}, A. 2012, \apjl, 758, L13, \dodoi{10.1088/2041-8205/758/1/L13}

\bibitem[{{Inoue} {et~al.}(2013){Inoue}, {Inoue}, {Kobayashi}, {Makiya}, {Niino}, \& {Totani}}]{2013Inoue}
{Inoue}, Y., {Inoue}, S., {Kobayashi}, M. A.~R., {et~al.} 2013, \apj, 768, 197, \dodoi{10.1088/0004-637X/768/2/197}

\bibitem[{{Isler} {et~al.}(2013){Isler}, {Urry}, {Coppi}, {Bailyn}, {Chatterjee}, {Fossati}, {Bonning}, {Maraschi}, \& {Buxton}}]{2013Isler}
{Isler}, J.~C., {Urry}, C.~M., {Coppi}, P., {et~al.} 2013, \apj, 779, 100, \dodoi{10.1088/0004-637X/779/2/100}

\bibitem[{{Jankowsky} {et~al.}(2022){Jankowsky}, {Ait-Benkhali}, {Zacharias}, \& {Wagner}}]{2022ATel15552....1J}
{Jankowsky}, F., {Ait-Benkhali}, F., {Zacharias}, M., \& {Wagner}, S.~J. 2022, The Astronomer's Telegram, 15552, 1

\bibitem[{{Kadler} {et~al.}(2022){Kadler}, {Stevens}, {Ojha}, {Edwards}, \& {Roesch}}]{2022ATel15536....1K}
{Kadler}, M., {Stevens}, J., {Ojha}, R., {Edwards}, P.~G., \& {Roesch}, F. 2022, The Astronomer's Telegram, 15536, 1

\bibitem[{{Kneiske} {et~al.}(2004){Kneiske}, {Bretz}, {Mannheim}, \& {Hartmann}}]{2004Kneiske}
{Kneiske}, T.~M., {Bretz}, T., {Mannheim}, K., \& {Hartmann}, D.~H. 2004, \aap, 413, 807, \dodoi{10.1051/0004-6361:20031542}

\bibitem[{{Kneiske} \& {Dole}(2010)}]{2010Kneiske}
{Kneiske}, T.~M., \& {Dole}, H. 2010, \aap, 515, A19, \dodoi{10.1051/0004-6361/200912000}

\bibitem[{{La Mura}(2022)}]{2022ATel15807....1L}
{La Mura}, G. 2022, The Astronomer's Telegram, 15807, 1

\bibitem[{Lei \& Wang(2014)}]{lei2014}
Lei, M., \& Wang, J. 2014, \pasj, 66, 7, \dodoi{10.1093/pasj/pst008}

\bibitem[{{Le{\'o}n-Tavares} {et~al.}(2013){Le{\'o}n-Tavares}, {Chavushyan}, {Pati{\~n}o-{\'A}lvarez}, {Valtaoja}, {Arshakian}, {Popovi{\'c}}, {Tornikoski}, {Lobanov}, {Carrami{\~n}ana}, {Carrasco}, \& {L{\"a}hteenm{\"a}ki}}]{2013Leon}
{Le{\'o}n-Tavares}, J., {Chavushyan}, V., {Pati{\~n}o-{\'A}lvarez}, V., {et~al.} 2013, \apjl, 763, L36, \dodoi{10.1088/2041-8205/763/2/L36}

\bibitem[{Liu \& Bai(2006)}]{Liu_2006}
Liu, H.~T., \& Bai, J.~M. 2006, The Astrophysical Journal, 653, 1089, \dodoi{10.1086/509097}

\bibitem[{{MAGIC Collaboration} {et~al.}(2018){MAGIC Collaboration}, {Acciari}, {Ansoldi}, {Antonelli}, {Arbet Engels}, {Arcaro}, {Baack}, {Babi{\'c}}, {Banerjee}, {Bangale}, {Barres de Almeida}, {Barrio}, {Bednarek}, {Bernardini}, {Berti}, {Besenrieder}, {Bhattacharyya}, {Bigongiari}, {Biland}, {Blanch}, {Bonnoli}, {Carosi}, {Ceribella}, {Cikota}, {Colak}, {Colin}, {Colombo}, {Contreras}, {Cortina}, {Covino}, {D'Elia}, {da Vela}, {Dazzi}, {de Angelis}, {de Lotto}, {Delfino}, {Delgado}, {di Pierro}, {Do Souto Espi{\~n}era}, {Dom{\'\i}nguez}, {Dominis Prester}, {Dorner}, {Doro}, {Einecke}, {Elsaesser}, {Fallah Ramazani}, {Fattorini}, {Fern{\'a}ndez-Barral}, {Ferrara}, {Fidalgo}, {Foffano}, {Fonseca}, {Font}, {Fruck}, {Galindo}, {Gallozzi}, {Garc{\'\i}a L{\'o}pez}, {Garczarczyk}, {Gaug}, {Giammaria}, {Godinovi{\'c}}, {Guberman}, {Hadasch}, {Hahn}, {Hassan}, {Herrera}, {Hoang}, {Hrupec}, {Inoue}, {Ishio}, {Iwamura}, {Kubo}, {Kushida}, {Kuve{\v{z}}di{\'c}}, {Lamastra}, {Lelas}, {Leone}, {Lindfors}, {Lombardi},
  {Longo}, {L{\'o}pez}, {L{\'o}pez-Oramas}, {Maggio}, {Majumdar}, {Makariev}, {Maneva}, {Manganaro}, {Mannheim}, {Maraschi}, {Mariotti}, {Mart{\'\i}nez}, {Masuda}, {Mazin}, {Minev}, {Miranda}, {Mirzoyan}, {Molina}, {Moralejo}, {Moreno}, {Moretti}, {Munar-Adrover}, {Neustroev}, {Niedzwiecki}, {Nievas Rosillo}, {Nigro}, {Nilsson}, {Ninci}, {Nishijima}, {Noda}, {Nogu{\'e}s}, {Paiano}, {Palacio}, {Paneque}, {Paoletti}, {Paredes}, {Pedaletti}, {Pe{\~n}il}, {Peresano}, {Persic}, {Prada Moroni}, {Prandini}, {Puljak}, {Garcia}, {Rhode}, {Rib{\'o}}, {Rico}, {Righi}, {Rugliancich}, {Saha}, {Saito}, {Satalecka}, {Schweizer}, {Sitarek}, {{\v{S}}nidari{\'c}}, {Sobczynska}, {Somero}, {Stamerra}, {Strzys}, {Suri{\'c}}, {Tavecchio}, {Temnikov}, {Terzi{\'c}}, {Teshima}, {Torres-Alb{\`a}}, {Tsujimoto}, {van Scherpenberg}, {Vanzo}, {Vazquez Acosta}, {Vovk}, {Ward}, {Will}, {Zari{\'c}}, {Fermi-Lat Collaboration}, {Becerra Gonz{\'a}lez}, {Raiteri}, {Sandrinelli}, {Hovatta}, {Kiehlmann}, {Max-Moerbeck}, {Tornikoski},
  {L{\"a}hteenm{\"a}ki}, {Tammi}, {Ramakrishnan}, {Thum}, {Agudo}, {Molina}, {G{\'o}mez}, {Fuentes}, {Casadio}, {Traianou}, {Myserlis}, \& {Kim}}]{2018A&A...619A.159M}
{MAGIC Collaboration}, {Acciari}, V.~A., {Ansoldi}, S., {et~al.} 2018, \aap, 619, A159, \dodoi{10.1051/0004-6361/201833618}

\bibitem[{Maraschi \& Tavecchio(2003)}]{Maraschi_2003}
Maraschi, L., \& Tavecchio, F. 2003, ApJ, 593, 667, \dodoi{10.1086/342118}

\bibitem[{Meyer {et~al.}(2019)Meyer, Scargle, \& Blandford}]{Meyer_2019}
Meyer, M., Scargle, J.~D., \& Blandford, R.~D. 2019, ApJ, 877, 39, \dodoi{10.3847/1538-4357/ab1651}

\bibitem[{{Nalewajko} {et~al.}(2012){Nalewajko}, {Begelman}, {Cerutti}, {Uzdensky}, \& {Sikora}}]{2012MNRAS.425.2519N}
{Nalewajko}, K., {Begelman}, M.~C., {Cerutti}, B., {Uzdensky}, D.~A., \& {Sikora}, M. 2012, \mnras, 425, 2519, \dodoi{10.1111/j.1365-2966.2012.21721.x}

\bibitem[{{Poutanen} \& {Stern}(2010)}]{Poutanen2010}
{Poutanen}, J., \& {Stern}, B. 2010, \apjl, 717, L118, \dodoi{10.1088/2041-8205/717/2/L118}

\bibitem[{{Rieger}(2019)}]{2019Galax...7...28R}
{Rieger}, F. 2019, Galaxies, 7, 28, \dodoi{10.3390/galaxies7010028}

\bibitem[{Sbarrato {et~al.}(2012)Sbarrato, Ghisellini, Maraschi, \& Colpi}]{10.1111/j.1365-2966.2012.20442.x}
Sbarrato, T., Ghisellini, G., Maraschi, L., \& Colpi, M. 2012, MNRAS, 421, 1764, \dodoi{10.1111/j.1365-2966.2012.20442.x}

\bibitem[{Scargle {et~al.}(2013)Scargle, Norris, Jackson, \& Chiang}]{Scargle_2013}
Scargle, J.~D., Norris, J.~P., Jackson, B., \& Chiang, J. 2013, The Astrophysical Journal, 764, 167, \dodoi{10.1088/0004-637X/764/2/167}

\bibitem[{{Scully} {et~al.}(2014){Scully}, {Malkan}, \& {Stecker}}]{2014Scully}
{Scully}, S.~T., {Malkan}, M.~A., \& {Stecker}, F.~W. 2014, \apj, 784, 138, \dodoi{10.1088/0004-637X/784/2/138}

\bibitem[{{Shukla} \& {Mannheim}(2020)}]{Shukla2020}
{Shukla}, A., \& {Mannheim}, K. 2020, Nature Communications, 11, 4176, \dodoi{10.1038/s41467-020-17912-z}

\bibitem[{{Sol} {et~al.}(2013){Sol}, {Zech}, {Boisson}, {Barres de Almeida}, {Biteau}, {Contreras}, {Giebels}, {Hassan}, {Inoue}, {Katarzy{\'n}ski}, {Krawczynski}, {Mirabal}, {Poutanen}, {Rieger}, {Totani}, {Benbow}, {Cerruti}, {Errando}, {Fallon}, {de Gouveia Dal Pino}, {Hinton}, {Inoue}, {Lenain}, {Neronov}, {Takahashi}, {Takami}, {White}, \& {CTA Consortium}}]{2013APh....43..215S}
{Sol}, H., {Zech}, A., {Boisson}, C., {et~al.} 2013, Astroparticle Physics, 43, 215, \dodoi{10.1016/j.astropartphys.2012.12.005}

\bibitem[{{Spada} {et~al.}(2001){Spada}, {Ghisellini}, {Lazzati}, \& {Celotti}}]{2001spada}
{Spada}, M., {Ghisellini}, G., {Lazzati}, D., \& {Celotti}, A. 2001, \mnras, 325, 1559, \dodoi{10.1046/j.1365-8711.2001.04557.x}

\bibitem[{{Stecker} {et~al.}(2006){Stecker}, {Malkan}, \& {Scully}}]{2006Stecker}
{Stecker}, F.~W., {Malkan}, M.~A., \& {Scully}, S.~T. 2006, \apj, 648, 774, \dodoi{10.1086/506188}

\bibitem[{{Stern} \& {Poutanen}(2014)}]{Stern_Poutanen_2014}
{Stern}, B.~E., \& {Poutanen}, J. 2014, \apj, 794, 8, \dodoi{10.1088/0004-637X/794/1/8}

\bibitem[{{Tanaka} {et~al.}(2011){Tanaka}, {Stawarz}, {Thompson}, {D'Ammando}, {Fegan}, {Lott}, {Wood}, {Cheung}, {Finke}, {Buson}, {Escande}, {Saito}, {Ohno}, {Takahashi}, {Donato}, {Chiang}, {Giroletti}, {Schinzel}, {Iafrate}, {Longo}, \& {Ciprini}}]{2011ApJ...733...19T}
{Tanaka}, Y.~T., {Stawarz}, {\L}., {Thompson}, D.~J., {et~al.} 2011, \apj, 733, 19, \dodoi{10.1088/0004-637X/733/1/19}

\bibitem[{{Tavecchio} \& {Ghisellini}(2012)}]{2012arXiv1209.2291T}
{Tavecchio}, F., \& {Ghisellini}, G. 2012, arXiv e-prints, arXiv:1209.2291, \dodoi{10.48550/arXiv.1209.2291}

\bibitem[{{Uttley} {et~al.}(2005){Uttley}, {McHardy}, \& {Vaughan}}]{2005MNRAS.359..345U}
{Uttley}, P., {McHardy}, I.~M., \& {Vaughan}, S. 2005, \mnras, 359, 345, \dodoi{10.1111/j.1365-2966.2005.08886.x}

\bibitem[{{Verrecchia} {et~al.}(2022){Verrecchia}, {Pittori}, {Piano}, {Tavani}, {Panebianco}, {Bulgarelli}, {Fioretti}, {Parmiggiani}, {Addis}, {Baroncelli}, {Di Piano}, {Lucarelli}, {Vercellone}, {Cardillo}, {Ursi}, {Casentini}, {Donnarumma}, {Gianotti}, {Trifoglio}, {Giuliani}, {Mereghetti}, {Caraveo}, {Perotti}, {Chen}, {Argan}, {Costa}, {Del Monte}, {Evangelista}, {Feroci}, {Foffano}, {Lapshov}, {Menegoni}, {Pacciani}, {Soffitta}, {Vittorini}, {Lazzarotto}, {Di Cocco}, {Fuschino}, {Galli}, {Labanti}, {Marisaldi}, {Pellizzoni}, {Pilia}, {Trois}, {Barbiellini}, {Longo}, {Vallazza}, {Morselli}, {Picozza}, {Prest}, {Lipari}, {Zanello}, {Cattaneo}, {Rappoldi}, {Ferrari}, {Antonelli}, {Giommi}, {Salotti}, {Valentini}, \& {D'Amico}}]{2022ATel15818....1V}
{Verrecchia}, F., {Pittori}, C., {Piano}, G., {et~al.} 2022, The Astronomer's Telegram, 15818, 1

\bibitem[{Wagner {et~al.}(2021)Wagner, Burd, Dorner, Mannheim, Buson, Gokus, Madejski, Scargle, Arbet-Engels, Baack, Balbo, Biland, Bretz, Buss, Elsaesser, Eisenberger, Hildebrand, Iotov, Kalenski, Neise, Noethe, Paravac, Rhode, Schleicher, Sliusar, \& Walter}]{Wagner:2021jn}
Wagner, S.~M., Burd, P., Dorner, D., {et~al.} 2021, in Proceedings of 37th International Cosmic Ray Conference {\textemdash} PoS(ICRC2021), Vol. 395, 868, \dodoi{10.22323/1.395.0868}

\bibitem[{Wood {et~al.}(2017)Wood, Caputo, Charles, Di~Mauro, Magill, \& Perkins}]{Wood:2017TJ}
Wood, M., Caputo, R., Charles, E., {et~al.} 2017, PoS, ICRC2017, 824, \dodoi{10.22323/1.301.0824}

\end{thebibliography}
\bibliographystyle{aasjournal}

\end{document}